\documentclass[12pt,titlepage]{article}
\usepackage{Transient}
\usepackage{multicol}
\usepackage[pdftex]{graphicx}
\usepackage{natbib}

\pdfoutput=1

\bibpunct[,]{(}{)}{;}{a}{}{,}

\begin{document}

\begin{titlepage}

\begin{center}
{\LARGE\bfseries
The Dynamic Radio Sky: \\
An Opportunity for Discovery
}
\end{center}

\vspace*{3ex} 

\begin{center}
{\large\bfseries
J.~Lazio${}^1$ (NRL), 
J.~S.~Bloom (UC Berkeley), 
G.~C.~Bower (UC Berkeley), 
J.~Cordes (Cornell, NAIC), 
S.~Croft (UC Berkeley), 
S.~Hyman (Sweet Briar), 
C.~Law (UC Berkeley), \& 
M.~McLaughlin~(WVU)
}
\end{center}

\begin{center}
Submitted to Astro2010: The Astronomy and Astrophysics Decadal Survey
\end{center}

\begin{center}
\includegraphics[width=0.99\textwidth]{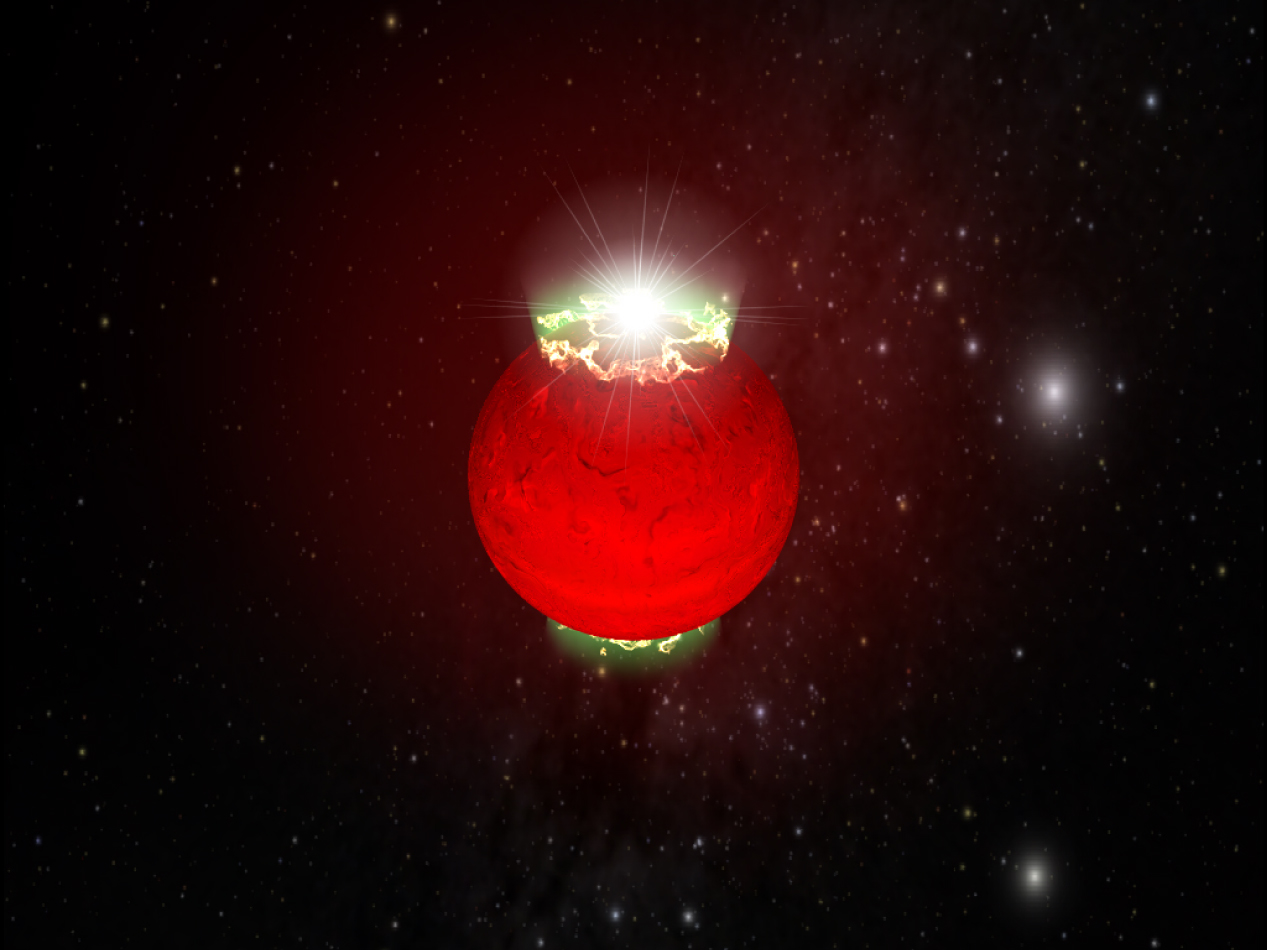}
\end{center}

\vfill

{\small
${}^1$ Contact information: 202-404-6329,
\textit{Joseph.Lazio@nrl.navy.mil}; Image credit: Hallinan et al.,
NRAO/AUI/NSF
}

\end{titlepage}

\begin{abstract}
\vspace*{-2ex}
The time domain of the sky has been only sparsely explored.
Nevertheless, recent discoveries from limited surveys and
serendipitous discoveries indicate that there is much to be found on
timescales from nanoseconds to years and at wavelengths from meters to
millimeters.  These observations have revealed unexpected phenomena
such as rotating radio transients and coherent pulses from brown
dwarfs.  Additionally, archival studies have found not-yet identified
radio transients without optical or high-energy hosts.  In addition
to the known classes of radio transients, possible other classes of
objects include extrapolations from known classes and exotica such as
orphan $\gamma$-ray burst afterglows, radio supernovae,
tidally-disrupted stars, flare stars, magnetars, and transmissions
from extraterrestrial civilizations.

Over the next decade, meter- and centimeter-wave radio telescopes with
improved sensitivity, wider fields of view, and flexible digital
signal processing will be able to explore radio transient parameter
space more comprehensively and systematically.
\end{abstract}

\section{Frontier Question: What New Sources and Phenomena Populate
the Sky?}\label{sec:question}

The available parameter space for transient surveys is extensive:
transients have been detected at, and are predicted for, all radio
wavelengths; timescales  range from
nanoseconds to the longest timescales probed; and transients may
originate from nearly all astrophysical environments including the
solar system, star-forming regions, the Galactic center, and other
galaxies.  

\emph{By observing the sky so as to preserve information about
the time domain, the past decade has illustrated that there is a
considerable potential for discovery.  Over the next decade, a
combination of increased sensitivity, field of view, and algorithmic
developments likely would yield transformational discoveries in
a wide range of astronomical fields.}

\section{Science Opportunity: The Dynamic Sky}\label{sec:dynamic}

Transient emission---bursts, flares, and pulses on time scales of less
than about~1~month---marks compact sources or the locations of
explosive or dynamic events.  Transient sources offer insight into a
variety of fundamental questions including
\setlength{\multicolsep}{4pt}
\setlength{\columnsep}{5pt}
\begin{multicols}{2}
\begin{itemize}
\setlength{\itemsep}{0pt}
\setlength{\parsep}{0pt}
\item Mechanisms of particle acceleration;
\item Possible physics beyond the Standard Model;
\item The physics of accretion and outflow;
\item Stellar evolution and death;
\item The nature of strong field gravity;
\item The nuclear equation of state;
\item The cosmological star formation history;
\item Probing the intervening medium(a); and
\item The possibility of extraterrestrial (ET) civilizations.
\end{itemize}
\end{multicols}

Much of astronomy's progress over the last half of the
$20^{\mathrm{th}}$ Century resulted from opening new spectral windows.
With essentially the entire spectrum having been explored at some
level, we must look to other parts of parameter space---such as
increased sensitivity, field of view, or the time domain---for future
transformational discoveries.

The time domain appears ripe for new exploration as
observations over the past decade have emphasized that the sky may be
quite dynamic---known sources have been discovered to behave in new ways and
what may be entirely new classes of sources have been discovered.
Radio observations triggered by high-energy observations
(e.g., observations of $\gamma$-ray burst [GRB] afterglows), monitoring
programs of known high-energy transients (e.g., radio monitoring of
X-ray binaries), giant pulses from the Crab pulsar, a small number of
dedicated radio transient surveys, and the serendipitous discovery of
transient radio sources (e.g., near the Galactic center, brown dwarfs)
all suggest that the sky is likely to be quite active on timescales from
nanoseconds to years and at wavelengths from meters to millimeters.

\section{Scientific Context: The Transient Sky}\label{sec:dynamic.highlight}

Classes of transients are diverse, ranging from nearby stars
to cosmological distances (GRBs), and touching upon nearly every
aspect of astronomy, astrophysics, and astrobiology.
Table~\ref{tab:drs.summary} lists a series of known, hypothesized, and
exotic classes of radio transients.  In the remainder of this section,
we provide two case studies and brief discussions of other classes of
transients.

\begin{table}
\begin{center}
\caption{Illustrations of Classes of Transients\label{tab:drs.summary}}
\begin{tabular}{p{0.3\textwidth}p{0.3\textwidth}p{0.3\textwidth}}
\hline\hline
\textbf{Known Classes}
	& \textbf{Extrapolations of Known Physics} 
	& \textbf{Exotica} \\
\hline
brown dwarfs, flare stars
	& extrasolar planets 
	& signals from ET civilizations \\
\hline
pulsar giant pulses, 
intermittant pulsars,
magnetar flares,
X-ray binaries      
	& giant pulses, flares from neutron stars in other galaxies 
	& electromagnetic counterparts to gravitational wave events \\
\hline
radio supernovae, 
GRB afterglows 
	& prompt emission from GRBs, orphan GRB afterglows 
	& annihilating black holes \\
\hline
variability from interstellar propagation
	& variability from intergalactic propagation 
	& \\
\hline\hline
\end{tabular}
\end{center}
\end{table}

\subsection{Case Study: Rotating Radio Transients---A New Population
of Neutron Stars}

The first pulsars were discovered through visual inspection of pen
chart recordings, which revealed the presence of individual radio
pulses spaced by the neutron star rotation period.  It was soon
realized that Fourier methods were far more sensitive to the periodic
emission believed to be characteristic of all radio pulsars, and
periodicity searches have been used in the discovery of over 1800
radio pulsars.

In~2003, the Parkes Multibeam Survey had covered the entire Galactic
plane visible from Parkes, finding over~700 new pulsars.  The data
were then re-analyzed for single, dispersed pulses, revealing a new population of neutron stars only detectable
through their individual radio bursts \citep{mll+06}.  
The average pulse rates of these 11 sources were (3~min)$^{-1}$
to (3~hr)$^{-1}$.  Periods ranging from~0.7--7~s were eventually inferred
from the differences between the pulse arrival times. These
periods are comparable to those of traditional radio pulsars, and
confirmed the neutron star nature of these sources, dubbed 
Rotating Radio Transients (RRATs).

Since the discovery of the original 11 RRATs, interest in single radio
pulse searches has increased dramatically.  Single pulse searches are
incorporated in the pipeline of current pulsar surveys, and a great
deal of archival pulsar search data has been reanalyzed. Currently,
roughly 30 RRATs are known, with this number increasing steadily.

What makes RRATs so different from normal pulsars, and how might they
be related to other classes of neutron stars? Perhaps fundamental
properties such as magnetic field or age contribute to the radio
sporadicity, or their emission could be due to external influences
such as a debris disk \citep{cs08}.  Another fundamental issue is the
total number of these sources.  Their sporadicity makes them difficult
to detect, and it is likely that the population of RRATs outnumbers
that of normal pulsars, leading \cite{kk08} to conclude that the
neutron star population is \emph{not} consistent with the Galactic
supernova rate.

In summary, 
the RRATs are an example of an unexpected source class
discovered through simple but new transient detection algorithms.

\subsection{Case Study: Unexplained Transient Events}\label{sec:unknown}

\begin{figure}[tb]
\begin{center}
\includegraphics[width=0.75\textwidth]{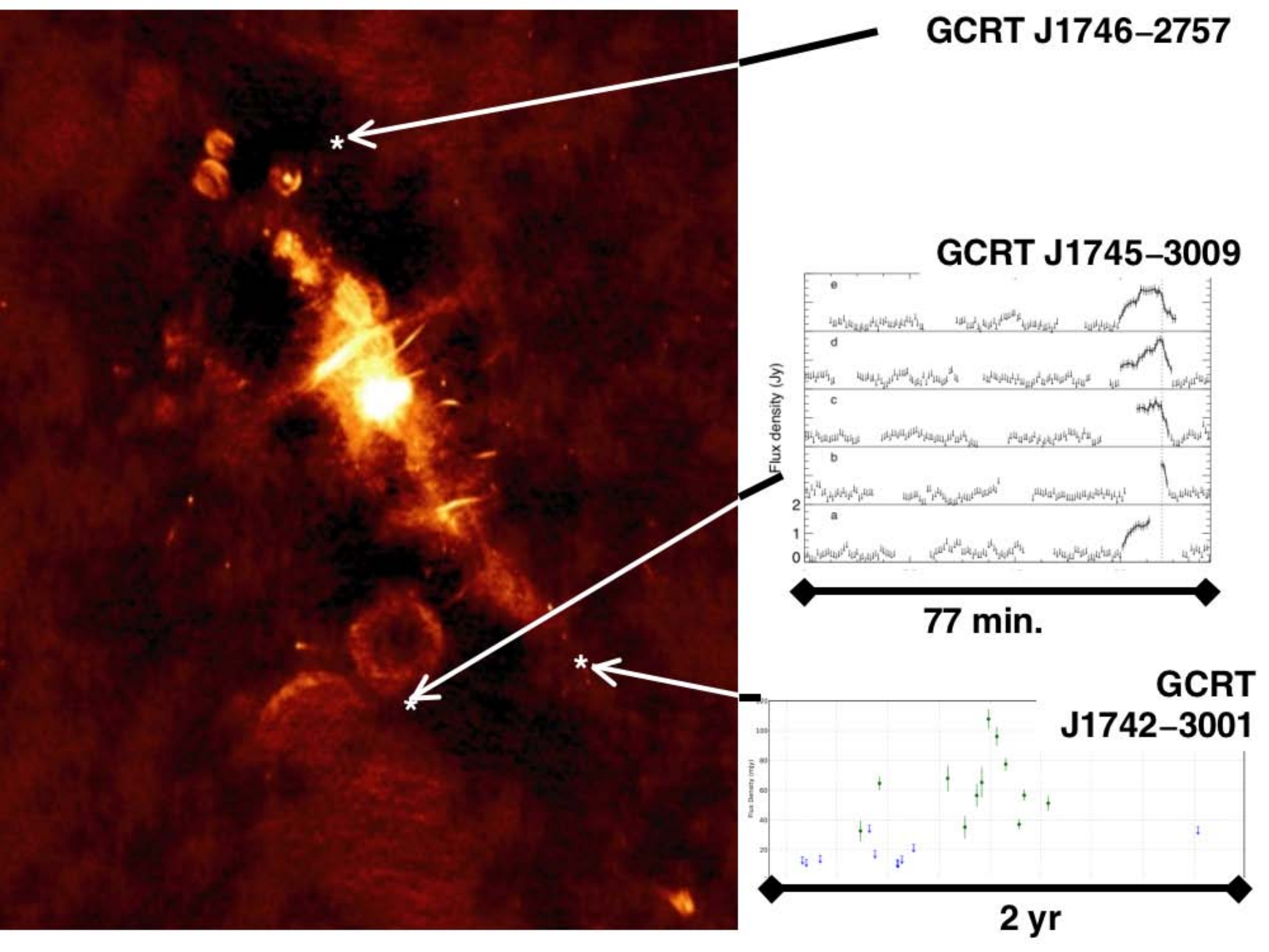}
\end{center}
\vspace*{-5ex}
\caption{%
Illustration of the diversity of the light curves for transients
toward the Galactic center \citep{hlkb02,hlkrmy-z05,hwlpskr09}.
The transient GCRT~J1745$-$3009 burst several
times (duration $\sim 10$~min.) during a 6-hr observation, with
subsequent bursts detected over the next 1.5~yr; GCRT~J1742$-$3001
brightened and faded over several months,   
preceded 6~months earlier by intermittent bursts; and GCRT~J1746$-$2757 was detected in only a single epoch.
None of these objects has
been identified nor has a multi-wavelength counterpart been found.
The background image is the Galactic center at~330~MHz, and the total
time devoted to the monitoring project, in both new and archival
observations, is about~150~hr.}
\label{fig:drs.gc}
\end{figure}

\begin{figure}[htb]
\vspace*{-1ex}
\begin{center}
\parbox[b]{0.67\textwidth}{%
\includegraphics[width=0.33\textwidth]{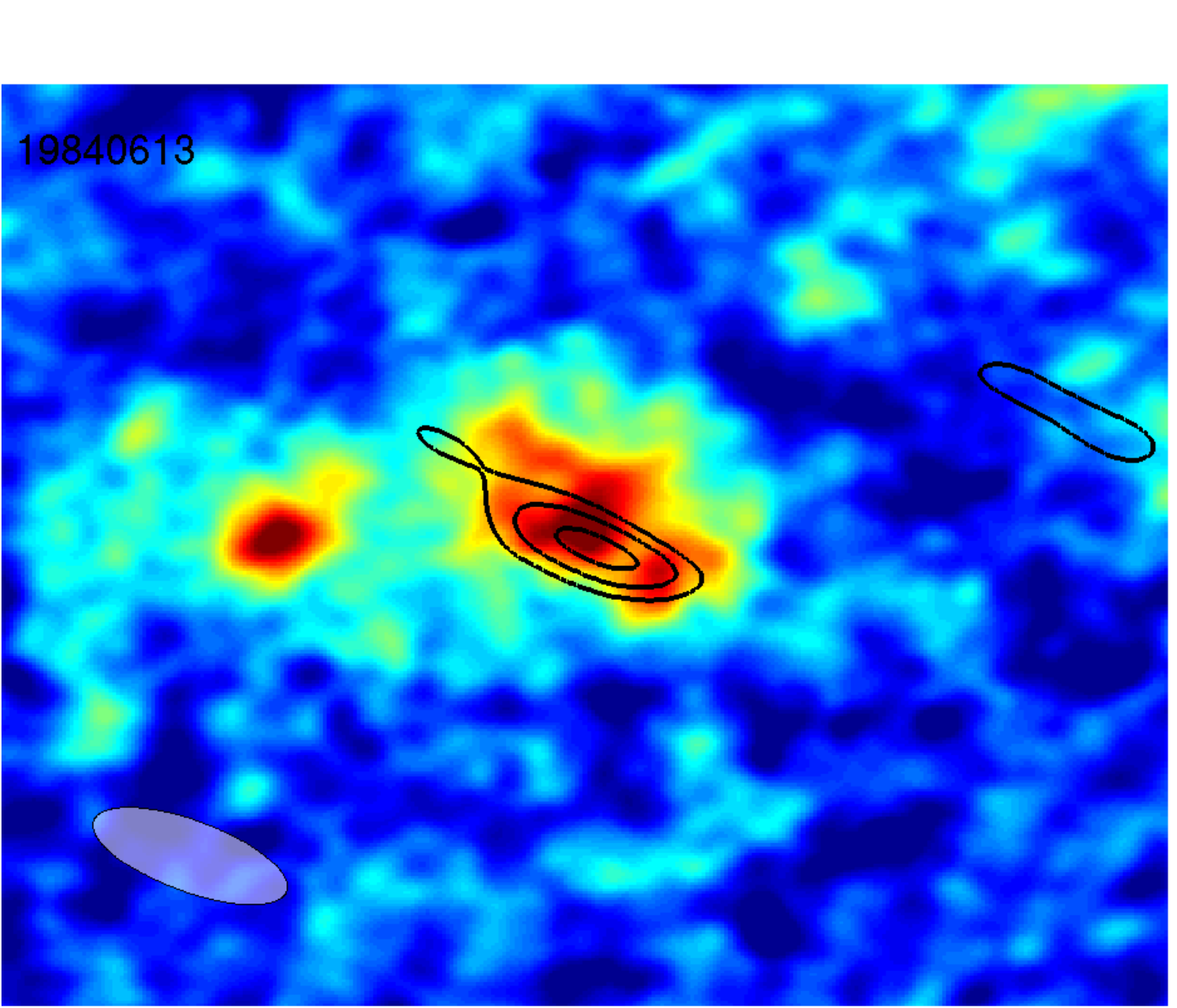}~\includegraphics[width=0.33\textwidth]{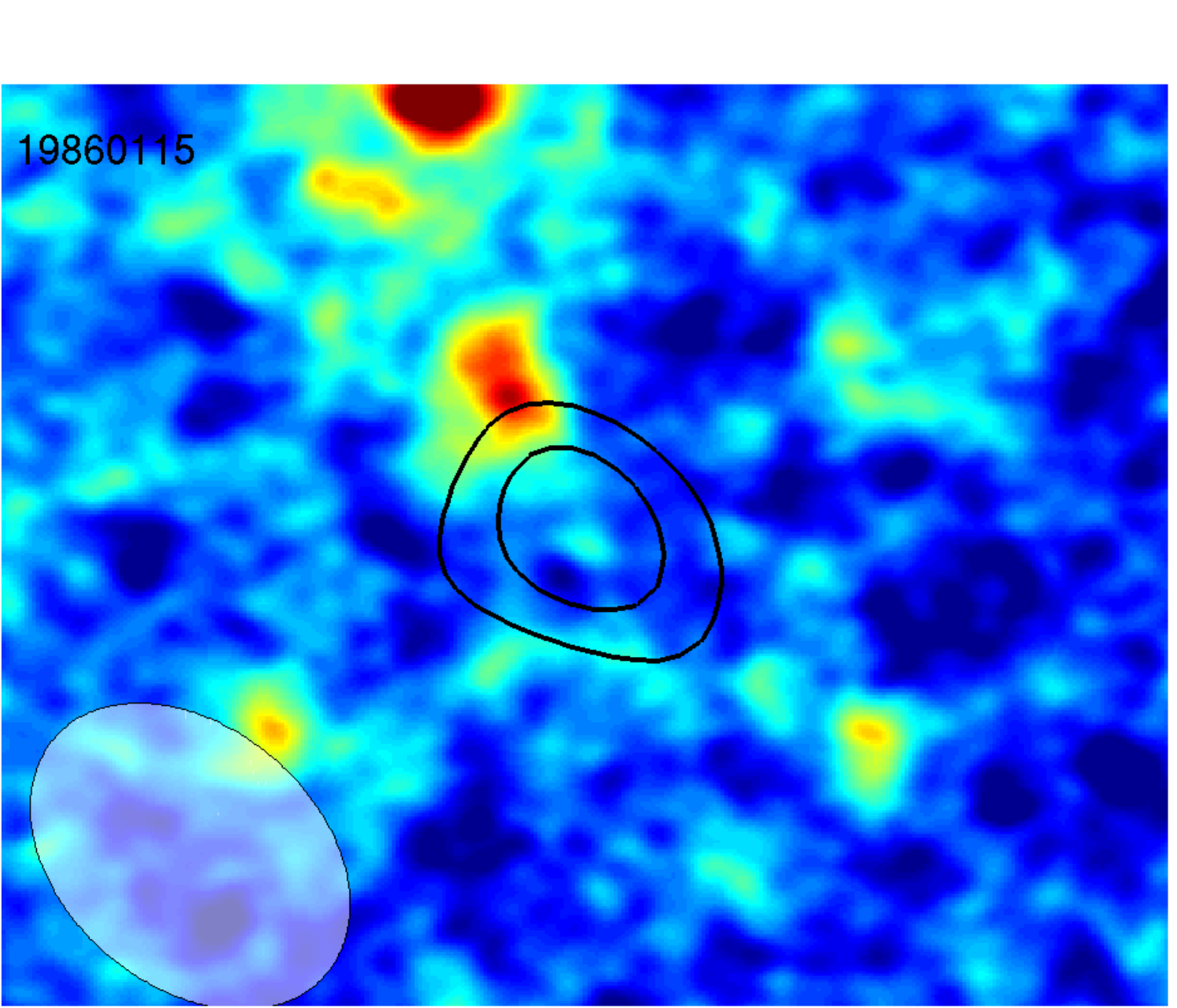}
\includegraphics[width=0.33\textwidth]{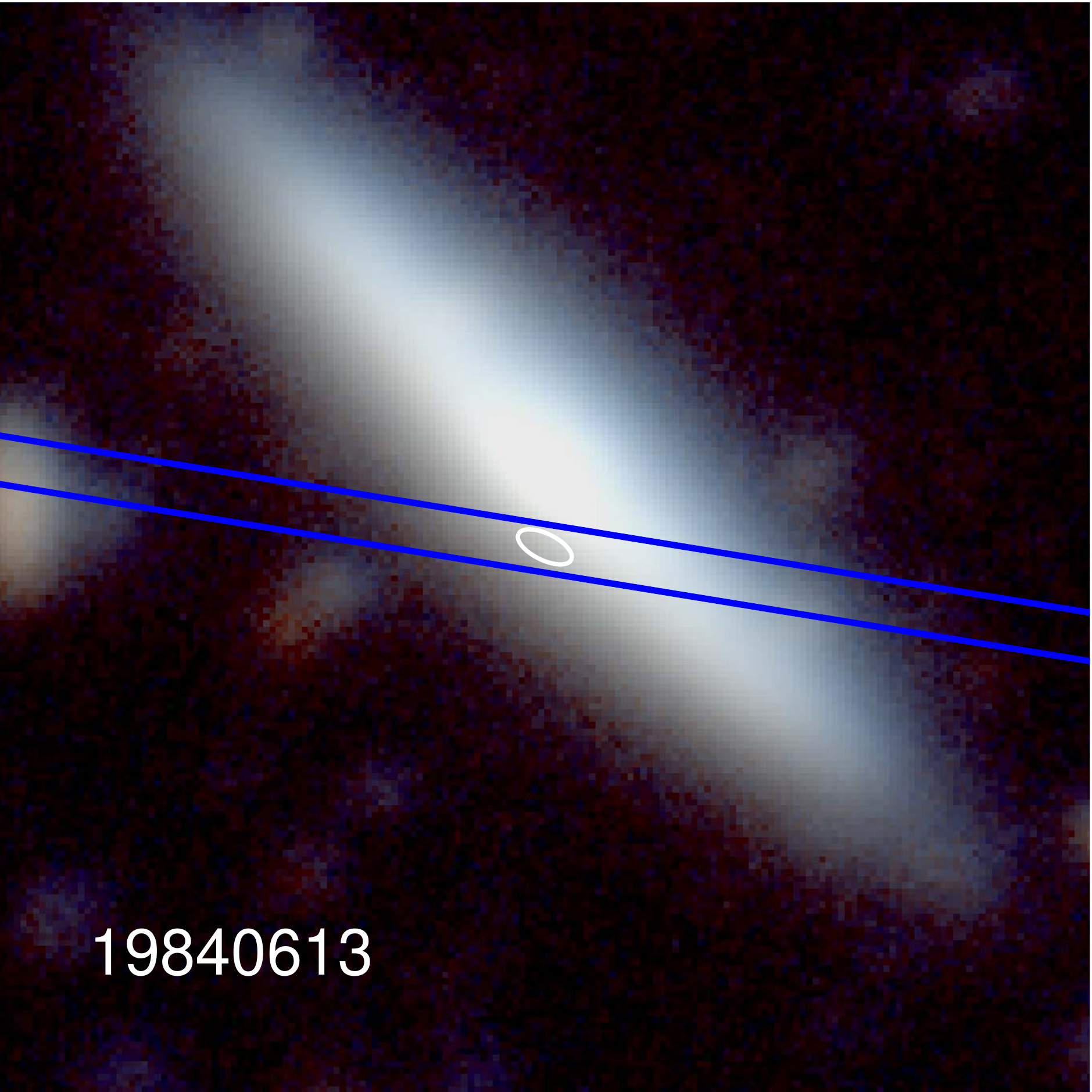}~\includegraphics[width=0.33\textwidth]{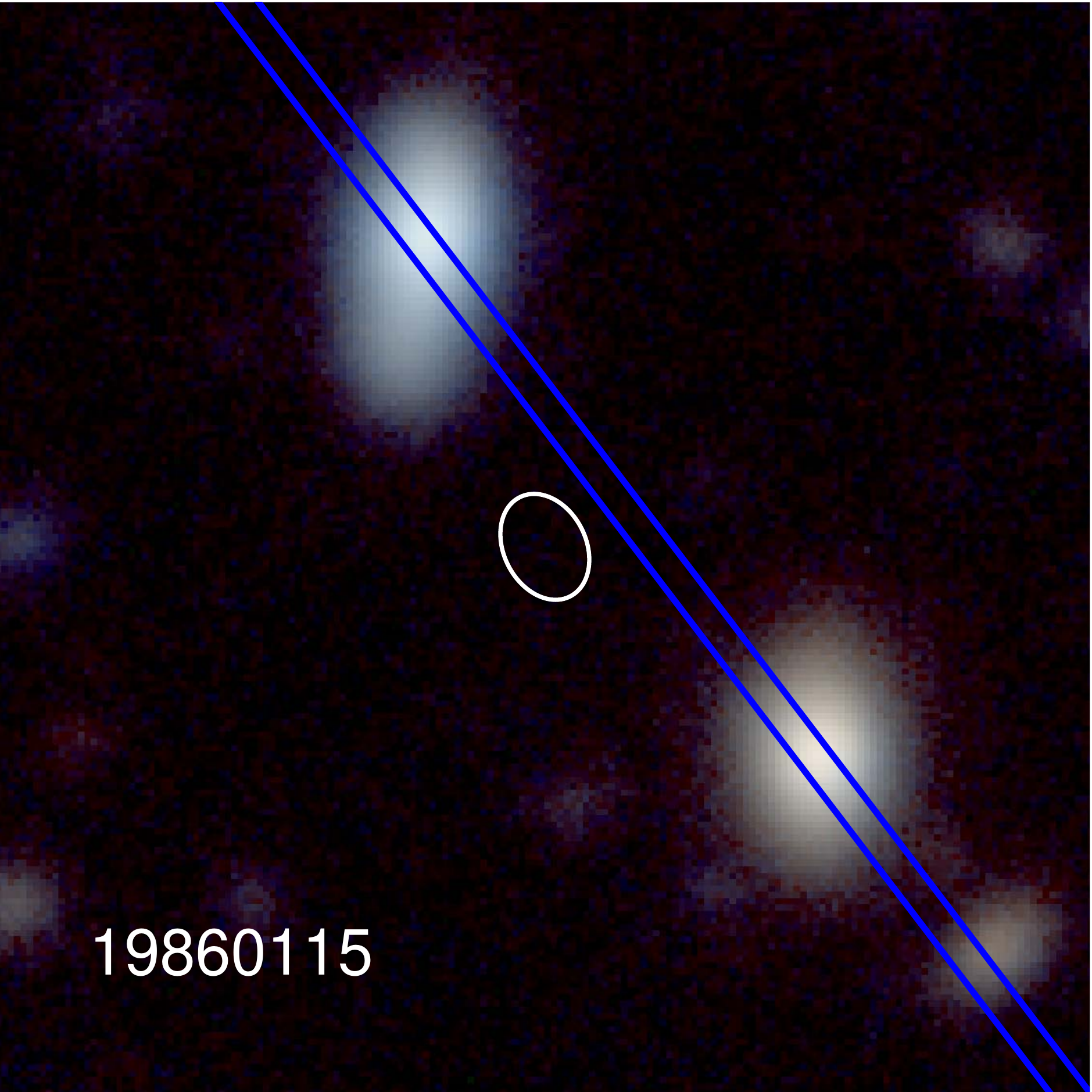}%
}~\parbox[b]{0.3\textwidth}{%
 \caption{%
Two radio transients found in a survey of~944 epochs of a
blank field from the VLA archives \citep{bsb+07}; there is no clear
object class identification for these or eight other transients.
(\textit{Top}) Contours indicate the transients' locations on the deep
radio image.  (\textit{Bottom}) The positions of the radio
transients overlaid on deep Keck~G and~R band images.  RT~19840613 is
offset by~3~kpc from the nucleus of a spiral galaxy at $z=0.04$;
RT~19860115 has no radio or optical counterpart.}
\label{fig:drs.bower}
}
\end{center}
\end{figure}

Figures~\ref{fig:drs.gc} and~\ref{fig:drs.bower} illustrate the
potential diversity of objects to be discovered.  These transients
were discovered in a combination of new and archival observations
toward the Galactic center (Figure~\ref{fig:drs.gc}) or in archival
observations of a ``blank field'' (Figure~\ref{fig:drs.bower}).
Archival data have proven particularly valuable resources for these
programs as both span 1--2~decades of time.  Most of the transients
shown in these figures have no multi-wavelength counterparts, nor are
they associated with any known transient classes.  Possible
explanations for the various transients range from rare, extremely
luminous flares from Galactic M dwarfs and brown dwarfs to GRB
afterglows.

\subsection{Diverse Populations: Opportunity for Discovery}\label{sec:other}

\setlength{\leftmargini}{0.75em}
\begin{description}
\setlength{\itemsep}{0pt}

\item[Flare Stars, Brown Dwarfs, and Extrasolar Planets:]%
Active stars and star systems have long been known to produce radio
flares attributed to particle acceleration from magnetic field
activity \citep{g02}.  More recently, flares from late-type stars (dM)
and brown dwarfs have been discovered \citep{bbb+01,hbl+07}, in some
cases with periodicities indicative of rotation.  The radio emission
from these late-type stellar objects is far stronger than expected
from the Benz-G\"udel relation for X-ray and radio emission from
main-sequence stars.  Finally, Jupiter is
radio bright below~40~MHz, and many stars with ``hot Jupiters'' show
signatures of magnetic star-planet interactions
\citep{swbgk05}, so extrasolar planets may also be radio sources
\citep{z07}.

\item[Pulsar Giant Pulses---Relativistic Magnetohydymamics and the
Intergalactic Medium:]
While all pulsars show pulse-to-pulse intensity variations, some pulsars  emit so-called
``giant'' pulses, with strengths 100 or even 1000 times the
mean pulse intensity. The Crab was the first pulsar
found to exhibit this phenomenon, and giant pulses have since been
detected from numerous other pulsars \citep{cstt96,rj01,jr03}.  
Pulses with flux densities of order $10^{3}$~Jy at 5~GHz and with durations of only
2~ns have been detected from the Crab \citep{hkwe03}. These
``nano-giant'' pulses imply brightness temperatures of
10$^{38}$~\hbox{K}, by far the most luminous emission from any
astronomical object.
In
addition to being probes of particle acceleration in the pulsar
magnetosphere, giant pulses may serve as probes of the local
intergalactic medium \citep{mc03}.

\item[Radio Supernovae and GRBs:]%
Observations of the kind possible with the new radio telescopes (i.e.,
frequent monitoring of large areas of sky) can be used to find those  
GRBs and supernovae that emit in the radio, as well as to follow up on  
such transients detected at other wavelengths. Multi-wavelength,  
multi-epoch observations \citep[e.g.,][]{c+06} can provide  
information on progenitors, the surrounding medium, and models of GRB  
energetics and beaming.  Of special interest is finding so-called ``orphan afterglows,'' those without $\gamma$-ray trigger. The demographics of orphan afterglows directly inform the geometry and hence energetics of the events \citep[e.g.,][]{lowg-y02}.

\item[Intraday Variability, AGN Central Engines, and Interstellar \& Intergalactic Media:]%
Intraday variability (IDV)---interstellar scintillation
of extremely compact components ($\sim 10$~$\mu$as) in \hbox{AGN}---occurs at frequencies near~5~GHz.  The typical modulation amplitude is
a few percent, but occasional sources display much larger modulations
\citep{k-cjwtr01,ljbk-cmrt03}; in
\emph{extreme scattering events},  modulations greater
than 50\% on time scales of days to months are obtained
\citep{fdjh87}.  The existence
of compact components in AGN may prove to be a sensitive probe of
their central engines, innermost regions of the jet, or both,
complementing $\gamma$-ray observations.  Finally, in order for AGN to
be sufficiently compact to scintillate, their signals must not have
been affected substantially by  propagation through the
\emph{intergalactic medium}.  Given that the dominant baryonic
component of the Universe is likely to be in a warm-hot intergalactic
medium, the presence of IDV can also constrain the
properties of the intergalactic medium.

\item[Annihilating Black Holes:]%
Annihilating black holes are predicted to produce radio bursts
\citep{r77}.  Advances in $\gamma$-ray detectors has renewed interest
in possible high-energy signatures from primordial black holes
\citep{dls02,lab+06}.  Observations at the extremes of the
electromagnetic spectrum are complementary as radio observations
attempt to detect the pulse from an individual primordial black hole,
while high-energy observations generally search for the integrated
emission.

\item[Gravitational Wave Events:]
The progenitors for gravitational wave events may generate
associated electromagnetic signals or pulses.  For example, the
in-spiral of a binary neutron star system, one of the key targets for
\hbox{LIGO}, may produce electromagnetic pulses, both at high
energies and in the radio due to the interaction of the magnetospheres of
the neutron stars \citep[e.g.,][]{hl01}.  
More generally, the combined detection of both electomagnetic and
gravitational wave signals may be required to produce 
localizations and understanding of the gravitational wave emitters
\citep{khm07}.
See also the whitepaper on the GW-EM connection \citep{bhh+09}.

\item[Extraterrestrial transmitters:]%
While none are known, searches for extraterrestrial intelligence
(SETI) have found non-repeating signals that are otherwise consistent
with the expected signal from an ET transmitter.  \cite{cls97} show
how ET signals could appear transient, even if intrinsically steady.
\end{description}

\section{Advancing the Science: Exploring Phase Space}\label{sec:dynamic.phase}

\emph{Over the next decade, great progress is possible in the study of
transients.  Specific steps include (1)~Explicit time-domain
processing of data coupled with algorithmic developments, particularly
in the area of identification and classification of transients; and~(2)~Exploitation of telescopes with higher sensitivities, wider
fields of view, or both.}

The transient detection figure of merit at radio wavelengths is
\begin{equation}
\mathrm{FoM}_t
 = \Omega \left(\frac{A_{\mathrm{eff}}}{T_{\mathrm{sys}}}\right)^2 K(\eta W, \tau W),
\label{eqn:ssfom}
\end{equation}
which is a function of the telescope
sensitivity~$A_{\mathrm{eff}}/T_{\mathrm{sys}}$, instantaneous solid
angle~$\Omega$, typical time duration of the
transient~$W$,  event rate~$\eta$, and the time per telescope
pointing (``dwell time'') $\tau$.  The function $K(\eta W,
\tau W)$ incorporates the likelihood of detecting a particular kind of
transient.
Roughly, one can separate transients surveys into two
classes:~(1)~Burst searches that probe timescales of less than
about~1~s for which $\Omega$ is large  but
$A_{\mathrm{eff}}/T_{\mathrm{sys}}$ is small; and~(2)~Imaging
surveys conducted with interferometers that typically probe timescales
of tens of seconds and longer and for which $\Omega$ is small but
$A_{\mathrm{eff}}/T_{\mathrm{sys}}$ is large.

\vspace*{-1ex}
\begin{enumerate}
\item \textbf{Explicit time-domain processing of data and algorithmic developments:}
Since the discovery of RRATs, interest in single radio
pulse searches has increased dramatically.  Searches for single,
dispersed pulses now are incorporated in the software pipeline of
current pulsar surveys, such as those at Arecibo, the \hbox{GBT}, and
Parkes, and archival pulsar data have been reanalyzed.  While
time-domain processing is not yet standard for many interferometers,
the \hbox{ASKAP}, \hbox{ATA},
\hbox{EVLA}, \hbox{LOFAR}, \hbox{LWA}, \hbox{MWA}, and eventually the
\hbox{SKA} offer new possibilities for expanding time-domain
processing to interferometric imaging.  Further, the interferometers
offer the possibility of much higher positional information for
transients, which is essential for multi-wavelength study.

A number of algorithmic improvements would yield improved use of the
existing telescopes and likely a higher yield from future telescopes.
\vspace*{-1ex}
\begin{packeditemize}
\item The vast storage and computational requirements of transient searches,
particularly in the case of imaging interferometers, requires the
development of near real-time transient analysis pipelines.  The
\hbox{ATA}, \hbox{LOFAR}, and MWA projects are all engaged in the
development of such first-generation pipelines.

\item The identification, avoidance, and excision of radio frequency
interference (RFI) produced by civil or military transmitters
operating in the radio spectrum is required.  These transmitters are
often orders of magnitude stronger than the desired astronomical
signal.

\item The identification and classification of transients is a
challenge that is broader than simply radio wavelength transients.
\end{packeditemize}

\item \textbf{Exploitation of telescopes with higher sensitivities, wider fields of view:}
Generally, both $A_{\mathrm{eff}}/T_{\mathrm{sys}}$ and
$\Omega$ should be large, though depending upon the class of transient
and its luminosity function (if known), it may be possible to trade
$A_{\mathrm{eff}}/T_{\mathrm{sys}}$ vs.\ $\Omega$.  For instance,
X- and $\gamma$-ray instruments with large solid angle coverage and
high time resolution have had great success in finding 
transients, even if the detectors were not particularly
sensitive.

In the last decade, the field of view of the Arecibo telescope
around~1~GHz was expanded by a factor of~7 with a new feed system
(ALFA).  In the next decade, additional field of view expansion
technologies such as \emph{phased-array feeds} offer the potential of
expanding the fields of view of single-dish telescopes such as Arecibo
and the \hbox{GBT} by factors of~10 or more.

For imaging surveys, \hbox{LOFAR}, the \hbox{LWA} and the MWA promise
much higher sensitivities at low radio frequencies for which the
fields of view are naturally large ($\sim 10$~deg.${}^2$).  The ASKAP
and ATA both offer the promise of much larger fields of view ($\sim
10$~deg.${}^2$) at frequencies near~1~GHz, while the EVLA will provide
a factor of~10 in sensitivity improvements across its entire
operational range (1--50~GHz).  All of these imaging
interferometers also can be \emph{sub-arrayed}, providing improvements in
field of view ($\sim 100$~deg.${}^2$), at the cost of sensitivity.
\end{enumerate}
\vspace*{-2ex}

Looking toward the next decade and to the era of the \hbox{SKA}, the
above advances in searches for transient radio sources promise to
transform our understanding of the dynamic Universe.

\vspace*{2ex}

\begin{multicols}{2}

\end{multicols}

\end{document}